# INFERENCE USING SHAPE-RESTRICTED REGRESSION SPLINES

By Mary C. Meyer

*Colorado State University*

Regression splines are smooth, flexible, and parsimonious non-parametric function estimators. They are known to be sensitive to knot number and placement, but if assumptions such as monotonicity or convexity may be imposed on the regression function, the shape-restricted regression splines are robust to knot choices. Monotone regression splines were introduced by Ramsay [*Statist. Sci.* **3** (1998) 425–461], but were limited to quadratic and lower order. In this paper an algorithm for the cubic monotone case is proposed, and the method is extended to convex constraints and variants such as increasing-concave. The restricted versions have smaller squared error loss than the unrestricted splines, although they have the same convergence rates. The relatively small degrees of freedom of the model and the insensitivity of the fits to the knot choices allow for practical inference methods; the computational efficiency allows for back-fitting of additive models. Tests of constant versus increasing and linear versus convex regression function, when implemented with shape-restricted regression splines, have higher power than the standard version using ordinary shape-restricted regression.

**1. Introduction.** We consider the regression model

$$(1) \qquad y_i = f(x_i) + \sigma \epsilon_i, \qquad i = 1, \ldots, n,$$

where the errors are i.i.d. Nonparametric regression methods provide estimates for $f$ using minimal assumptions, and are appropriate when a parametric form is unavailable. Many methods assume only some sort of smoothness; three of the most widely used of these are the kernel smoother, the smoothing spline, and regression splines. These require user-specified choices as bandwidth, or smoothing parameter, or number and placement of knots. If the fits are not robust to these choices, inference about the regression function is problematic.

Fits using only shape restrictions such as monotonicity or convexity do not require user-defined parameters, but are typically not smooth, nor are the fits parsimonious, in that the model degrees of freedom is in some sense large. Some inference methods have been developed: for testing constant versus increasing regression function [see Robertson, Wright and Dyskstra (1988), hereafter RWD) and for linear versus convex regression function [see Meyer (2003)], test statistics with exact distributions under the null hypothesis and normal errors assumption have been derived.

Several methods combining smoothing and shape restrictions have been proposed. Mammen (1991) investigated the asymptotic behavior of the monotonized kernel estimator, and alternatively, the kernel-smoothed monotone regression estimator, showing that either obtains the $n^{-2/5}$ pointwise convergence rate of the original kernel estimator.

The constrained smoothing splines are more challenging to obtain. The expression

$$(2) \qquad \sum_{i=1}^{n} [y_i - f(x_i)]^2 + \lambda \int_a^b [f^{(k)}(x)]^2 \, dx$$

is to be minimized over the set of $f$ satisfying the constraints. The constant $\lambda$ is called the "smoothing parameter" because larger values result in smoother fits. Tantiyaswasdikul and Woodroofe (1994) characterized the monotone smoothing splines for $k = 1$. The natural cubic smoothing spline minimizes the expression (2) for $k = 2$, but although it is relatively easy to impose shape restrictions at the observed $x$-values, ensuring that the restrictions hold between the observations is more difficult. For monotone cubic interpolation, it is known that additional knots must be placed between the observations [Fredenhagen, Oberle and Opfer (1999)], but the exact placement of these knots is not understood. Delecroix, Simioni and Thomas-Agnan (1995) demonstrated through simulations that imposing shape restrictions on top of smoothing leads to substantial reductions of squared error loss for moderately sized samples and many choices of underlying function, error variance, and shape. The monotone smoothing splines were shown by Mammen and Thomas-Agnan (1999) to obtain the optimal $n^{-p/(2p+1)}$ convergence rate, where $p$ is the maximum of $k$ and the order of the polynomial.

In this paper the estimation of $f$ using shape-restricted regression splines is discussed, and inference methods proposed. Regression splines are simple and straight-forward; a set of basis functions are provided which act as the regressors in an ordinary least-squares model. The basis functions are smooth, often piecewise polynomials of degree $d$ between the user-specified knots, with $d - 1$ continuous derivatives. The optimal convergence rate is attained; Huang and Stone (2002) provide a nice discussion of the trade-off between estimation error and approximation error in choosing the number of knots. The set of linear combinations of the basis functions typically provides more than enough flexibility to fit a scatterplot; in fact, in the absense of shape restrictions, the flexibility for a rather small number of knots is so great that most of the regression spline literature concerns guarding against the over-fitting of data. Friedman and Silverman (1989) discussed knot placement of unconstrained polynomial splines, based on forward-backward model selection. They imposed a minimum span for possible knot locations, based on the idea that the smoother will "follow runs" of sequential positive or negative errors, and result in over-fitting. Eilers and Marx (1996) proposed using a rather large number of knots, but including a penalty term to reduce the flexibility of the regression estimator and guard against over-fitting.

These techniques are not necessary for the shape-restricted regression spline. Typically, when assumptions about both shape and smoothness are warranted, the fits are robust to the choice of smoothing parameters. The monotonicity constraints, for example, do not allow peaks and valleys in the fit, although rounded "steps" may be present, and convexity (or concavity) constraints disallow any sort of wiggling. In other words, the shape restrictions themselves provide some smoothing, and obviate cures for over-fitting.

An example data set involving age and income data is shown in Figure 1. These data were used as an example in Ruppert, Wand and Carroll (2003) (hereafter RWC) and represent a sample of 205 Canadian workers, all of whom were educated to grade 13. The relationship between log(income) and age is to be modeled nonparametrically. In plot (a) there are three examples of penalized regression splines. The solid line represents the cross-validation choice of the smoothing parameter; the dashed line has a larger parameter, and the dot-dash line has a smaller. Smaller penalty parameters allow the fit to be steeper at the left-hand side, but allow for wiggles toward the right-hand side. The cross-validation choice still has a dip at about age 41–42. If we believe that the true relationship between log(income) and age should not have a dip, we might assume the relationship is convex. In plot (b) we see the cubic regression splines, constrained to be convex, for three choices of interior knots. The fits are very close to one another, almost indistinguishable, and have no trouble with the steep slope at the left.

A second motivating example uses the "onion" dataset from RWC. When onions are planted more densely in a field, it is expected that the yield per plant will decrease, and we suppose here that the log of the yield is also convex in the density of the planting. The smoothing splines presented in plot (a) of Figure 2 are constrained to be decreasing and convex (at the observed $x$-values), for three choices of the smoothing parameter. The solid curve represents the cross-validation choice. The parameter for the dotted curve is larger, and the fit is close to a straight line. The smoothing parameter is smaller for the dashed curve, which rises more sharply at the left. The decreasing convex cubic regression splines are shown in (b), for two, four, and six interior knots. Again, these fits lie almost on top of each other.

Robustness of nonparametric function estimator to user inputs is important for inference, to ensure that different choices will not produce different answers. The rest of the paper is organized as follows. The algorithm for the computation of shape-restricted regression splines is described in Section 2. Hypothesis tests for constant versus increasing and linear versus convex re-gression function are discussed in Section 3, where an exact test statistic is obtained. Degrees of freedom and estimation of model variance are discussed in Section 4.

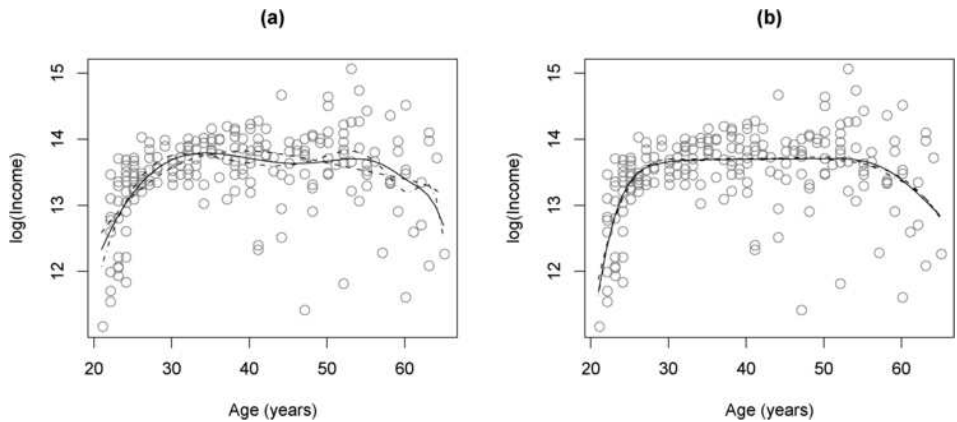

Fig. 1. *Age and income data for a sample of Canadians. (a) Penalized cubic splines for three values of the penalty parameter; the solid curve represents the cross-validation choice. (b) Cubic regression splines constrained to be concave, with three (dashed), five (solid) and seven (dot-dash) interior knots.*

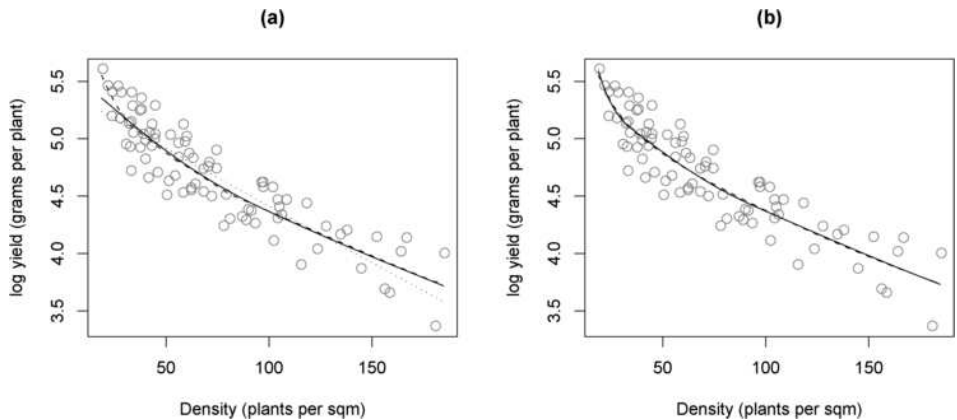

Fig. 2. *Comparison of fits to the onion data. (a) The natural cubic smoothing spline fits constrained to be decreasing and convex, for three choices of smoothing parameter; the solid line is the cross-validation choice. (b) The decreasing convex cubic regression spline, for three choices of knots.*

**2. Computing the estimator.** Ramsay (1988) introduced monotone regression splines to fit a scatterplot of data $(x_i, y_i)$, for $i = 1, \ldots, n$, where the $x_i$ are ordered. For regression splines of order $k$, choose $l$ grid points $t_{k+1}, \ldots, t_{k+l}$, and define knots $x_1 = t_1 = \cdots = t_k < \cdots < t_{l+k+1} = \cdots = t_{l+2k} = x_n$. Then the number of $M$-spline basis functions is $m = l + k$; these are given recursively as follows. Order 1 $M$-splines are the piecewise constant (step functions)

$$M_i^{(1)}(x) = \begin{cases} \dfrac{1}{t_{i+1} - t_i}, & \text{for } t_i \leq x \leq t_{i+1}, \\ 0, & \text{otherwise,} \end{cases}$$

for $i = 1, \ldots, l + 1$. Order $k$ $M$-splines are computed from the lower orders:

$$M_i^{(k)}(x) = \begin{cases} \dfrac{k[(x - t_i)M_i^{(k-1)}(x) + (t_{i+k} - x)M_{i+1}^{(k-1)}(x)]}{(k-1)(t_{i+k} - t_i)}, \\ \qquad \text{for } t_i \leq x \leq t_{i+k}, \\ 0, \quad \text{otherwise.} \end{cases}$$

Finally, the $I$-splines are

$$(3) \quad I_i^{(k)}(x) = \int_{t_1}^x M_i^{(k)}(u)\,du \qquad \text{for } i = 1, \ldots, l + k = m, \text{ for } x \in [x_1, x_n].$$

The regression function is estimated by a linear combination of the basis functions and the constant function. To constrain the estimator to be monotone, the coefficients of the basis functions must be nonnegative (the coefficient of the constant function is not constrained).

For convex $C$-splines are integrated, to get basis functions that are both increasing and convex. In particular,

$$(4) \quad C_i^{(k)}(x) = \int_{t_1}^x I_i^{(k)}(u)\,du \qquad \text{for } i = 1, \ldots, l + k = m \text{ for } x \in [x_1, x_n].$$

A convex regression function is estimated using linear combinations of the basis functions with nonnegative coefficients, plus an unrestricted linear combination of the constant function and the identity function $g(x) = x$. If the underlying regression function is *both* increasing and convex, we restrict the coefficient on the identity function also to be nonnegative.

Define the set of vectors $\boldsymbol{\sigma}^j$ in $\mathbb{R}^n$ containing the values of the $j$th basis function, evaluated at the $x$-values. For $k$th-order monotone regression splines, let $\sigma_i^j = I_j^{(k)}(x_i)$, for $j = 1, \ldots, m$, and $i = 1, \ldots, n$. For the convex case, simply substitute the $C$-spline basis functions for the $I_j$. Let $V$ be the linear space contained in the constraint set; for the monotone and monotone-convex cases, $V = \mathcal{L}(\mathbf{1})$, and for convex constraints, $V = \mathcal{L}(\mathbf{1}, \boldsymbol{x})$, where $\mathcal{L}$ denotes "linear space spanned by," $\mathbf{1} = (1, \ldots, 1)'$, and $\boldsymbol{x} = (x_1, \ldots, x_n)'$.

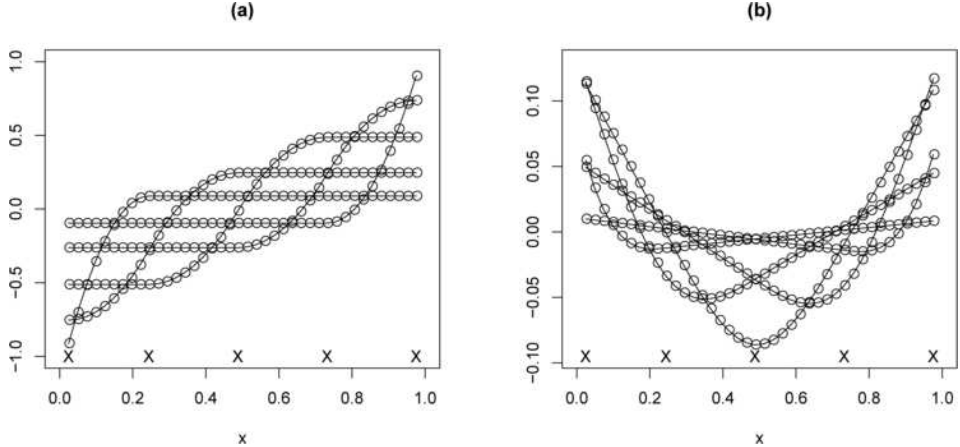

Fig. 3. *Spline basis functions (curves) using three equally spaced interior knots, along with edge vectors (with values marked as dots) for a data set with $n = 40$ equally spaced design points. The knot positions are marked with "X." (a) Quadratic monotone, scaled so that the vectors have mean zero. (b) Cubic convex, scaled so that the vectors are orthogonal to both the one-vector and the x-vector.*

A set of "generating vectors" that are orthogonal to $V$ are

$$\boldsymbol{\delta}^j = \boldsymbol{\sigma}^j - \Pi(\boldsymbol{\sigma}^j | V),$$

where $\Pi$ is a projection operator. Because the $\boldsymbol{\sigma}$ and $\boldsymbol{v}$ vectors form a linearly independent set, then the $\boldsymbol{\delta}$ vectors are also linearly independent. The constraint set $\mathcal{C}$ can then be characterized by

$$\mathcal{C} = \left\{ \boldsymbol{\theta} : \boldsymbol{\theta} = \boldsymbol{v} + \sum_{j=1}^{m} b_j \boldsymbol{\delta}^j, \text{ where } b_j \geq 0, j = 1, \ldots, m, \text{ and } \boldsymbol{v} \in V \right\}.$$

For $\boldsymbol{y} = \boldsymbol{\theta} + \boldsymbol{\epsilon}$, the least-squares estimator $\hat{\boldsymbol{\theta}}$ minimizes $\|\boldsymbol{y} - \boldsymbol{\theta}\|^2$ over $\boldsymbol{\theta} \in \mathcal{C}$, where the notation is $\|\boldsymbol{a}\|^2 = \sum a_i^2$. The set of nonnegative linear combinations of the $\boldsymbol{\delta}$ vectors is called the "constraint cone" $\Omega$, and $\hat{\boldsymbol{\theta}}$ can be found by projecting onto $\Omega$ and $V$ separately, then adding the projections. A depiction of the cone edges for piecewise quadratic increasing constraints using three interior knots and $n = 40$ is shown in Figure 3, plot (a). The piecewise cubic edge vectors for the convex case are shown in plot (b). The knots are marked on the plots with "X." The basis functions and basis vectors are easily modified for use with concave constraints, or increasing concave, and so on.

The third-order (quadratic) $I$-splines defined in (3) are "proper" set basis functions, in that the coefficients of the piecewise quadratic monotone basis functions being nonnegative is a necessary and sufficient condition for the linear combination to be nondecreasing. This can be seen by observing that, at each knot, exactly one basis function has a positive first derivative. Similarly, the first and second order $I$-splines form proper sets, but the cubic $I$-splines do not. A linear combination of cubic $I$-splines might be nondecreasing while one or more of the coefficients is negative, so that the least-squares estimator might lie outside of $\mathcal{C}$. The lack of a cubic version of the monotone splines is a drawback when the user would like a smooth first derivative. An approach to obtaining a solution is presented at the end of this section.

For the convex cubic splines defined in (3), we observe that, at each knot, exactly one basis function has a positive second derivative, so that the $C$-splines form proper sets of basis functions up to order four.

Ramsay (1988) provides a gradient-based algorithm for finding the least-squares solution for the $I$-splines. This converges to the solution in "infinitely many" steps, meaning that the algorithm produces a sequence of estimates that get closer to the true least-squares estimate, then stops when the gradient satisfies a convergence criterion. The algorithm proposed here finds the true solution in a small number of steps, by taking advantage of the fact that the constraint set is a closed convex polyhedral cone in $\mathbb{R}^n$.

Because $\mathcal{C}$ is convex, the minimizer $\hat{\boldsymbol{\theta}}$ of $\|\boldsymbol{y} - \boldsymbol{\theta}\|^2$ over $\boldsymbol{\theta} \in \mathcal{C}$ is unique, and the necessary and sufficient condition is $\langle \boldsymbol{y} - \hat{\boldsymbol{\theta}}, \hat{\boldsymbol{\theta}} - \boldsymbol{\theta} \rangle \geq \boldsymbol{0}$, for all $\boldsymbol{\theta} \in \mathcal{C}$ (RWD). Here the notation $\langle \boldsymbol{a}, \boldsymbol{b} \rangle$ denotes the inner product $\sum a_i b_i$. Because $\mathcal{C}$ is a cone, the condition may be written as

$$\langle \boldsymbol{y} - \hat{\boldsymbol{\theta}}, \hat{\boldsymbol{\theta}} \rangle = \boldsymbol{0} \tag{5}$$

and

$$\langle \boldsymbol{y} - \hat{\boldsymbol{\theta}}, \boldsymbol{\theta} \rangle \leq \boldsymbol{0} \qquad \text{for all } \boldsymbol{\theta} \in \mathcal{C}. \tag{6}$$

Subsets of the generators $\boldsymbol{\delta}^j$ define "faces" of the constraint cone. Any $J \subseteq \{1, 2, \ldots, m\}$ defines a set $\mathcal{F}(J) = \sum_{j \in J} b_j \boldsymbol{\delta}^j$, where $b_j > 0$ for $j \in J$. The constraint cone itself is a face with $J = \{1, \ldots, m\}$, and the origin is a face with $J$ equal to the empty set, so that there are $2^m$ faces. The projection onto the cone $\Omega$ will land on one of these faces, and in fact is a projection onto a linear subspace; for a proof of the following proposition, see Meyer (1999).

PROPOSITION 1. *Let $J$ be the subset of $\{1, \ldots, m\}$ such that the unique minimizer of $\|\boldsymbol{y} - \boldsymbol{\theta}\|^2$ is $\hat{\boldsymbol{\theta}} = \boldsymbol{v} + \sum_{j \in J} b_j \boldsymbol{\delta}^j$, where the $b_j$ are strictly positive for $j \in J$. Then $\hat{\boldsymbol{\theta}}$ is the projection of $\boldsymbol{y}$ onto the linear space spanned by vectors in $V$ and the $\boldsymbol{\delta}^j$, $j \in J$.*

The mixed primal-dual bases algorithm by Fraser and Massam (1989) and the hinge algorithm by Meyer (1996) are efficient ways to determine the set $J$. In either scheme, an initial set $J_0$ is proposed, and ordinary least squares projection onto the face with edges indexed by $J$ is determined. Edges are added or removed one-by-one (the rules are different for the two algorithms) until the correct $J$ [as determined by (1) and (2)] is found. A QR decomposition of the "design matrix" allows the new projection to be obtained from the old projection with less computation. To demonstrate the speed of the algorithm, 10,000 datasets of size $n = 100$, simulated using $y_i = x_i^2 + \epsilon_i$, with $x$ values equally spaced on $(0, 2)$ and i.i.d. standard normal errors. A piecewise quadratic monotone spline was fit with the hinge algorithm to each dataset, using four interior knots. At most ten iterations of the algorithm were needed; the modal number of iterations was five. When $n$ is increased to 500 and using six interior knots, the maximum number of iterations (out of 10,000 simulated data sets) was twelve, with a mode of seven.

*Consistency and rates of convergence.* The following proposition states that the shape-restricted version of the regression spline has smaller squared error loss than the unrestricted version, when the true regression function satisfies the shape assumptions. Therefore, the shape-restricted regression spline is consistent under the same regularity conditions as the ordinary piecewise polynomial regression spline, and the rate is at least as good. The proof is straight-forward.

PROPOSITION 2. *Let $\tilde{\mathcal{S}}$ be the linear space spanned by the $\boldsymbol{\delta}$ vectors and the vectors in $V$. Let $\theta_i = f(x_i)$, and assume the shape restrictions hold for $f$. Let $\hat{\boldsymbol{y}}$ be the unconstrained projection of $\boldsymbol{y}$ onto $\tilde{\mathcal{S}}$, and recall that $\hat{\boldsymbol{\theta}}$ is the projection of $\boldsymbol{y}$ onto $\mathcal{C}$. Then*

$$\|\hat{\boldsymbol{\theta}} - \boldsymbol{\theta}\| \leq \|\hat{\boldsymbol{y}} - \boldsymbol{\theta}\|,$$

*with equality only if $\hat{\boldsymbol{\theta}} = \hat{\boldsymbol{y}}$.*

Hwang and Stone (2002) show that for the unrestricted case and "bounded mesh ratio," the asymptotically optimal number of knots is $l \approx n^{1/(2p+1)}$, where $p$ is the order of the polynomial pieces. This choice allows the estimator to attain the pointwise convergence rate $O_p(n^{-p/(2p+1)})$.

*Choice of knots.* Shape restrictions impose some degree of smoothness by themselves. Combining shape restrictions with smoothing will result in estimators that are typically more robust than smoothing only, especially if a convexity assumption is warranted. The asymptotically optimal $n^{-1/(2p+1)}$ rounds to two or three for $n$ up to about 500, but if the underlying regression function shows a lot of variation such as rapid rises, more knots might be needed to follow the data adequately. This is demonstrated in the following simulations, which give some insight into knot choices.

TABLE 1

*Comparison of the root mean squared error loss for penalized splines with quadratic regression splines constrained to be increasing. Data were simulated using the indicated $f$, $\sigma^2 = 1$, and x-values equally spaced on $(0,1)$. PSPL indicates the penalized regression spline with $k = n/3$ knots, and cross-validation selection of the smoothing parameter; MR is ordinary monotone regression, MSPL2 is quadratic monotone regression spline with two interior knots, MSPL4 is defined similarly*

| | $f(x) = 4x$ | | | | $f(x) = 5\exp(10x-5)/[1+\exp(10x-5)]$ | | | |
|---|---|---|---|---|---|---|---|---|
| $n$ | PSPL | MR | MSPL2 | MSPL4 | PSPL | MR | MSPL2 | MSPL4 |
| 40 | 0.36 | 0.43 | 0.31 | 0.34 | 0.41 | 0.52 | 0.47 | 0.35 |
| 80 | 0.26 | 0.34 | 0.23 | 0.26 | 0.33 | 0.41 | 0.41 | 0.25 |
| 200 | 0.17 | 0.25 | 0.15 | 0.17 | 0.22 | 0.27 | 0.26 | 0.16 |

*Goodness of fit.* We compare the root mean squared error loss of the constrained regression splines with that of the penalized regression spline and the standard shape-restricted regression estimators using simulations. The squared error loss for the estimate $\hat{f}_j$ from the $j$th simulated dataset is

$$SEL_j = \frac{1}{n}\sum_{i=1}^{n}[\hat{f}_j(x_i) - f(x_i)]^2,$$

and the square roots of the average *SEL* are reported for 10,000 simulated datasets.

For monotone constraints, we use two underlying functions: $f(x) = 4x$ and $f(x) = 5\exp(10x-5)/(1+\exp(10x-5))$ over $(0,1)$, both with unit model variance. The first varies steadily, but the latter has a steep increase in the middle of the range of the data, which makes it difficult for the quadratic monotone spline if there are too few knots. The fit with two interior knots is not flexible enough to fit the steep rise adequately, although using four interior knots gives a substantial improvement. In summary, the 4-knot monotone spline does best for the sigmoidal data for all three sample sizes, followed by the penalized spline. The 2-knot spline does not do better than the unsmoothed monotone regression for this choice of function. For the linear function, the 2-knot spline performs best, followed by the 4-knot spline and the penalized spline, which out-perform the unsmoothed monotone regression.

The advantage of using monotone convex constraints when available is illustrated using the regression functions $f(x) = 4x$ and $f(x) = 4x^2$. In each case, the smoothed constrained fits do best, followed by the unsmoothed, constrained regression. The unconstrained fit is last in each case.

*Weighted regression.* Suppose we have the model $\boldsymbol{y} = \boldsymbol{\theta} + \sigma\boldsymbol{\epsilon}$, where $\text{cov}(\epsilon) = A$ for a known positive definite $A$. The weighted shape-restricted regression spline fit is easy to obtain with a transformation of the data and of the constraint cone. We transform the model to the i.i.d. case by multiplying the model through by $A^{-1/2}$, the inverse of the Cholesky decomposition of the covariance matrix, to get $\tilde{\boldsymbol{y}} = \tilde{\boldsymbol{\theta}} + \sigma\tilde{\boldsymbol{\epsilon}}$, and similarly transforming $\tilde{\boldsymbol{v}}^j = A^{-1/2}\boldsymbol{v}$, $j = 1, \ldots, r$ and $\tilde{\boldsymbol{\delta}}^j = A^{-1/2}\boldsymbol{\delta}^j$. The projection is performed using the transformed model, then the answer is subjected to the reverse transformation.

This is useful when there are multiple observations at some $x$-values. For example, the ages in the Canadian age and income dataset are integers ranging from 21 to 65, with one to twelve observations at each age. Because we want to have distinct $x$-values, we average the incomes for each age group and weight the regression according to the number of observations comprising the average. The fits shown in Figure 1(b) were computed using the weighted model. In Figure 4 we see the weighted residuals plotted against age for the fit with five interior knots. There is not so much of a "fanning out" pattern as might be expected from the scatterplot, partly because more observations were taken for the younger ages. In fact, if the outlier at age 57 is ignored, there is hardly any pattern to the residuals.

*Additive models.* The computational efficiency facilitates "back-fitting" additive models of the form

$$\boldsymbol{y} = \boldsymbol{\theta} + \boldsymbol{D}\boldsymbol{\beta} + \boldsymbol{\epsilon},$$

for an $n \times q$ design matrix $\boldsymbol{D}$ and $r$-dimensional parameter vector $\boldsymbol{\beta}$. For a given $\boldsymbol{\beta}$, $\hat{\boldsymbol{\theta}} \in \mathcal{C}$ may be found to minimize $\|(\boldsymbol{y} - \boldsymbol{D}\boldsymbol{\beta}) - \boldsymbol{\theta}\|^2$, using the projection algorithm. The space $\mathbb{R}^q$ is searched for the optimal $\boldsymbol{\beta}$.

A simple example is the two-parallel curves model. To illustrate, we use the onion data set again, with two locations of farms indicated as shown in Figure 5. The decreasing convex parallel cubic regression splines are shown for three choices of knots.

TABLE 2
*Same as for Table 1 except MCR is ordinary monotone-convex regression, MCSPL2 is cubic monotone-convex regression spline with two interior knots, MCSPL4 is defined similarly*

| $n$ | $f(x) = 4x$ | | | | $f(x) = 4x^2$ | | | |
|---|---|---|---|---|---|---|---|---|
| | **PSPL** | **MCR** | **MCSPL2** | **MCSPL4** | **PSPL** | **MCR** | **MCSPL2** | **MCSPL4** |
| 40 | 0.36 | 0.30 | 0.21 | 0.29 | 0.36 | 0.31 | 0.27 | 0.27 |
| 80 | 0.26 | 0.22 | 0.19 | 0.20 | 0.25 | 0.23 | 0.20 | 0.21 |
| 200 | 0.16 | 0.14 | 0.12 | 0.13 | 0.17 | 0.16 | 0.14 | 0.14 |

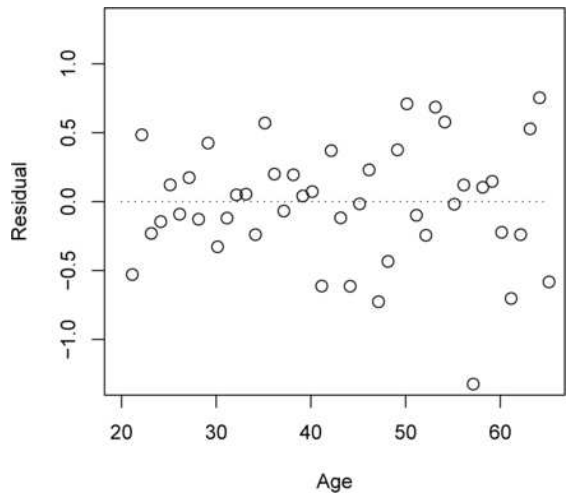

Fig. 4. *Weighted residuals for the fit to the Canadian age and income data.*

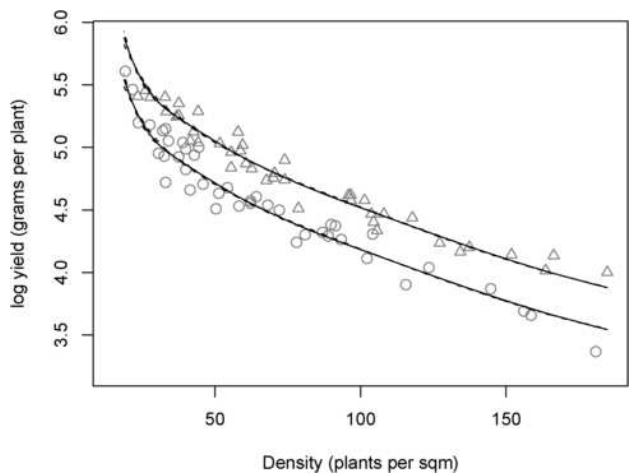

Fig. 5. *Log (yield) of onion plants against density of planting, for two locations. The Pumong Landing data are shown as triangles and the Virginia data are indicated with circles. The cubic regression splines constrained to be decreasing and convex are shown for two, four, and six equally spaced interior knots.*

All distinct $x$-values are used to define the spline basis functions. Given a choice of $\beta$, the least-squares parallel curves can be computed though a cone projection. The golden section search method [see Press et al. (1992), Section 10.1] is used to find the parameter $\beta$ representing the location effect, or the distance between the curves, that minimizes the overall sum of squared residuals. The estimates for $\beta$ do not vary substantially with choice of knots.

For two, three, and four interior knots, we estimate that the Pumong Landing log (yield) is 0.335 points higher than for the Virginia onions, and for five and six interior knots, the estimated difference is 0.338.

*Toward a monotone piecewise cubic regression spline.*   The necessary and sufficient conditions that a cubic function be monotone on an interval can not be written as a set of linear inequality constraints, so a proper set of basis functions for the monotone cubic regression splines does not exist. Fritsch and Carlson (1980) outline the conditions and provide an algorithm for the monotone piecewise cubic interpolation to monotone data, which will be part of the method proposed here.

The set of nondecreasing vectors in $\mathbb{R}^n$ form a polyhedral convex cone that can be characterized by $\Omega_M = \{\boldsymbol{\theta} : A\boldsymbol{\theta} \geq 0\}$, where the nonzero elements of the $(n-1) \times n$ matrix $A$ are $A_{i,i} = -1$ and $A_{i,i+1} = 1$, $i = 1, \ldots, n-1$. The cubic $M$-spline vectors span a linear subspace of $\mathbb{R}^n$, and the intersection of this subspace with $\Omega_M$ forms a polyhedral convex subcone $\Omega$. The edges of this cone can be used as basis vectors for the monotone cubic splines. This constrains the fit to be increasing at the observations, that is, $\hat{\theta}_1 \leq \cdots \leq \hat{\theta}_n$, but between observations the piecewise cubic function may be decreasing. Hence, we propose to use these basis vectors to get the least-squares solution $\hat{\boldsymbol{\theta}}$, then, if necessary, find the monotone interpolating cubic spline of Fritsch and Carlson (1980) to get $\hat{f}$. This interpolation does not minimize the integral of the squared second derivative of $f$, and, in fact, there is not a unique monotone cubic interpolation, but the optimal convergence rate is attained for any monotone interpolation.

The general method for finding the edges for the intersection of a convex cone with a linear subspace is as follows. Suppose a basis for the linear subspace is represented by the columns of the $n \times k$ matrix $V$, and the cone is defined by the $m \times n$ constraint matrix $A$, that is, $A\boldsymbol{\theta} \geq \boldsymbol{0}$. Then the vector $\boldsymbol{v} = V\boldsymbol{c}$ is in the intersection if $AV\boldsymbol{c} \geq 0$, so that the constraint matrix for the coefficient vector is the $m \times k$ matrix $AV$. Suppose the row space of $AV$ has dimension $q \leq k$. The edges of the cone are found as follows. If $\boldsymbol{r}_1, \ldots, \boldsymbol{r}_{q-1}$ are linearly independent rows of $AV$, $\boldsymbol{r}_0 \in \mathbb{R}^q$ is orthogonal to the space spanned by $\boldsymbol{r}_1, \ldots, \boldsymbol{r}_{q-1}$, and $AV\boldsymbol{r}_0 \geq \boldsymbol{0}$, then $\boldsymbol{r}_0$ is an edge. Furthermore, all edges are of this form. The proof of this claim can be found in Meyer (1999).

Using this method, we can find all linear combinations of the $M$-splines that are increasing at the observations. The number of edge vectors for the sub-cone can be quite large, which is typical for the case of more constraints than dimensions [see Meyer (1999) for more examples]. For three interior knots and $n = 40$, sixteen of the 233 edges are shown in Figure 6. The dots are the values of the edge vectors, and the lines are the piecewise cubic spline functions. The functions appear to be increasing, but some are not.

For example, the top left curve is increasing and then decreasing between the last two observation points. The cone formed by these edges is larger than that of the cubic $I$-splines, and any vector $\boldsymbol{\theta}$ that is a linear combination of the cubic $M$-spline basis vectors and satisfies $\theta_i \leq \theta_{i+1}$, $i = 1, \ldots, n-1$, is in the cone. Therefore, a monotone cubic spline interpolation of the projection of $\boldsymbol{y}$ onto the cone coincides with the least-squares solution and attains the optimal convergence rate. The piecewise cubic monotone spline is compared with the piecewise quadratic in Figure 7, for the same sigmoidal function used in the goodness of fit simulations. The cubic version is more flexible with fewer knots, and is able to follow the steeper rise in the data.

**3. Hypothesis testing.** One practical use for nonparametric function estimation is in testing proposed parametric models. The simplest examples are the test $H_0: f \equiv c$ versus $H_a: f$ is increasing, and the test of linear versus convex regression function. When the alternative fit is the ordinary (unsmoothed) shape-restricted estimator, a likelihood ratio approach yields a test statistic distributed as a mixture of beta random variables. See RWD, Chapter 2, for the monotone case and Meyer (2003) for the convex case. If we include the "smooth" assumption, and use the shape-restricted regression spline for the alternative fit, we get again a test statistic with a mixture of betas distribution. To show this, some characterizations of the constraint and polar cones are needed.

Let $S$ be the linear space spanned by the vectors $\boldsymbol{\delta}^j$, $j = 1, \ldots, m$. Then the polar cone is defined as all vectors in $S$ making obtuse angles with all vectors in $\Omega$:

$$\Omega^0 = \{\boldsymbol{\rho} \in S : \langle \boldsymbol{\rho}, \boldsymbol{\theta} \rangle \leq \boldsymbol{0}, \text{ for all } \boldsymbol{\theta} \in \Omega\}.$$

Let $\boldsymbol{\gamma}^1, \ldots, \boldsymbol{\gamma}^m$ be the rows of the matrix $[-(\Delta'\Delta)^{-1}\Delta']$, where the columns of the $n \times m$ matrix $\Delta$ are the $\boldsymbol{\delta}^j$. Then $\langle \boldsymbol{\gamma}^j, \boldsymbol{\delta}^i \rangle = 0$ for $i \neq j$, and $\langle \boldsymbol{\gamma}^j, \boldsymbol{\delta}^j \rangle = -1$, and by construction, $\boldsymbol{\gamma}^j \in \Omega^0$ and the $\boldsymbol{\gamma}^j$ span the same space as the $\boldsymbol{\delta}^j$ vectors. The proof of the next result is straight-forward.

PROPOSITION 3. *The vectors $\boldsymbol{\gamma}^j$ are generators of $\Omega^0$, that is,*

$$\Omega^0 = \left\{ \boldsymbol{\rho} : \boldsymbol{\rho} = \sum_{i=1}^m b_j \boldsymbol{\gamma}^j, \text{ where } b_j \geq 0, \text{ for all } j = 1, \ldots, m \right\}.$$

Let $r$ be the dimension of $V$, so that $r = 1$ for monotone or monotone-convex, and $r = 2$ for convex constraints. We may define $n - m - r$ linearly independent vectors $\boldsymbol{w}$ that are orthogonal to the $\boldsymbol{\delta}$ vectors and to $V$. The proof of the next proposition is given in the Appendix.

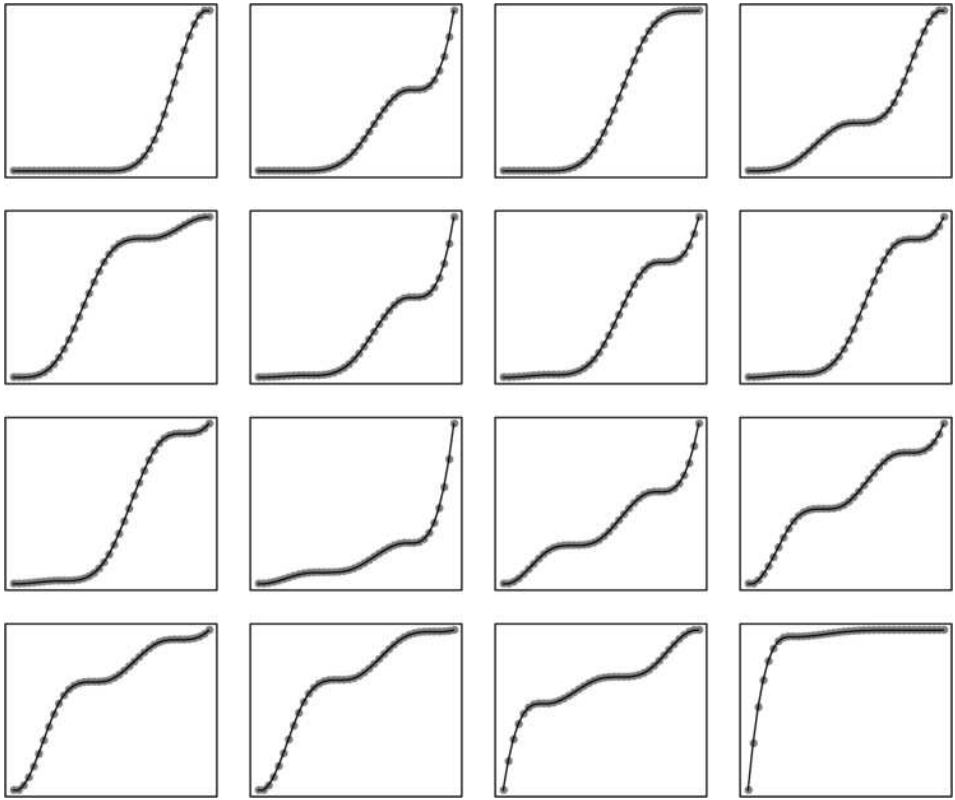

Fig. 6.  *Some of the edge vectors for the piecewise cubic regression spline, monotone at the x-observations, for three interior knots and n = 40.*

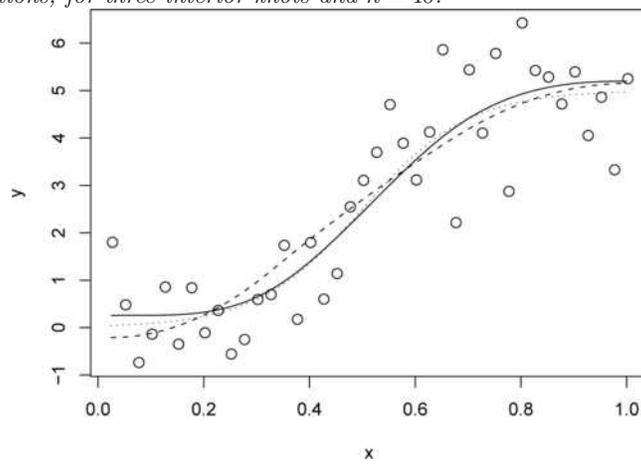

Fig. 7.  *Comparing the piecewise quadratic monotone spline (dashed) with the piecewise cubic (solid), both using two interior knots and n = 40. The true function is the dotted curve.*

PROPOSITION 4. *For any $\boldsymbol{y} \in \mathbb{R}^n$, there is a unique $J \subseteq \{1, \ldots, m\}$ so that $\boldsymbol{y}$ can be written uniquely as*

$$
(7) \qquad \boldsymbol{y} = \boldsymbol{v} + \sum_{j \in J} b_j \boldsymbol{\delta}^j + \sum_{j \notin J} b_j \boldsymbol{\gamma}^j + \sum_{j=1}^{n-m-r} c_j \boldsymbol{w}^j,
$$

*where $\boldsymbol{v} \in V$, $b_j > 0$ for $j \in J$ and $b_j \geq 0$ for $j \notin J$.*

Let $\chi_{01}^2 = (SSE_0 - SSE_1)/\sigma^2$, where $SSE_0$ is the residual sum of squares under the hypothesis $\boldsymbol{\theta} \in V$ and $SSE_1$ is that under $H_a : \boldsymbol{\theta} \in \mathcal{C}$. To derive the distribution of $\chi_{01}^2$ under $H_0$, we need to define "sectors" $C_J$, for all $J \subseteq \{1, \ldots, m\}$. We have

$$
C_J = \left\{ \boldsymbol{y} : \boldsymbol{y} = \boldsymbol{v} + \sum_{j \in J} b_j \boldsymbol{\delta}^j + \sum_{j \notin J} b_j \boldsymbol{\gamma}^j + \sum_{j=1}^{n-m-r} c_j \boldsymbol{w}^j \right\},
$$

where $\boldsymbol{v} \in V$, $b_j > 0$ for $j \in J$ and $b_j \geq 0$ for $j \notin J$. By Proposition 4, these $2^m$ sectors are disjoint and cover $\mathbb{R}^n$. Note that the interior of the constraint set is itself a sector with $J = \{1, \ldots, m\}$. We write

$$
P(\chi_{01}^2 \leq a) = \sum_{\text{subsets } J} P(\chi_{01}^2 \leq a | \boldsymbol{y} \in C_J) P(\boldsymbol{y} \in C_J).
$$

The derivation of Theorem 1 of Meyer (2003) applies, to show that under $H_0$, $P(\chi_{01}^2 \leq a | \boldsymbol{y} \in C_J) = P(\chi^2(n - d - r) \leq a)$, where $\chi^2(k)$ is a chi-square random variable with $k$ degrees of freedom, and $d$ is the number of indices in $J$. Therefore,

$$
P(\chi_{01}^2 \leq a) = \sum_{d=0}^{m} P(\chi^2(n - d - r) \leq a) P(D = d),
$$

where $D$ is a random variable indicating the number of $\boldsymbol{\delta}$ generators corresponding to the sector in which the data vector $\boldsymbol{y}$ falls, that is, the size of the set $J$.

The mixing distribution parameters $P(D = d)$ are found numerically as described in Meyer (2003). If the model variance $\sigma^2$ is not known, the test statistic

$$
B_{01} = \frac{\chi_{01}^2}{\chi_{01}^2 + SSE_1/\sigma^2} = \frac{SSE_0 - SSE_1}{SSE_0}
$$

has under $H_0$ a mixture of beta densities

$$
P(B_{01} \leq a) = \sum_{d=0}^{m} P\left[ B\left( \frac{d}{2}, \frac{n - d - r}{2} \right) \leq a \right] P(D = d),
$$

where $B(p, q)$ is a beta random variable with parameters $p$ and $q$. Because the mixing distribution can be determined as precisely as desired, the distribution of the test statistic under the null hypothesis is known exactly.

*Onion example.* The example of Figure 2 involves estimating yield as a function of planting density for onions. Suppose the farmer would prefer to use the simple linear relationship. Is there enough evidence to show that this is incorrect? The test for linear versus convex regression function provides a $p$-value of 0.0047 for two interior knots, $p = 0.0036$ for three, and $p = 0.0037$ for four.

*Simulations.* The power for the tests is compared with tests using the standard shape-restricted regression estimators, and also with the $F$-test. To compare tests of monotone versus increasing regression function, we choose two underlying regression functions (linear and "ramp"), three sample sizes, and for each sample size, three model standard deviations. The "ramp" regression function is $f(x) = \exp[8(x - 1/2)]$, so that it is flat at the left and increasing steeply at the right. For each sample size, the three model standard deviations were chosen so that the power of the $F$-test was 0.25, 0.50 and 0.75. The alternative fit for the $F$-test is a line, so that the model is not correct for the data generated from the ramp function. The idea is that practitioners might not know the true regression function, and might use the simplest choice of parametric function for the alternative fit in a "test for trend."

For each combination of sample size, model standard deviation, and underlying regression function, 10,000 datasets were simulated. For each dataset, three hypothesis tests were performed: the $F$-test for constant versus linear regression function, for constant versus increasing regression function using the monotone quadratic (IQRS), and the test for constant versus increasing regression function using the standard monotone regression estimator (MREG).

For the linear regression function, the $F$-test is the gold standard. We see that the tests using nonparametric alternative fits have smaller power, but the regression spline versions have power that is considerably closer to that of the $F$-test. For the ramp regression function, the power for the nonparametric tests is larger, and again the regression spline version has higher power than the standard version.

The next set of simulations compares the power for the linear versus convex regression function tests. The $F$-test uses the quadratic as the alternative fit, the "CCRS B-test" uses the convex piecewise cubic regression spline as the alternative fit, and "CREG B-test" uses the standard convex regression estimator. The two underlying regression functions are the quadratic and the "ramp" functions. Again, results are reported for combinations of underlying regression function, three sample sizes and three model standard deviations. For the quadratic underlying regression function, the $F$-test has highest power, and the power for the test using regression splines has higher

Table 3

*Power comparisons for the test of constant vs. monotone regression function. For the regression spline, the number of interior knots is $l = 2$, 2, and 3 corresponding to $n = 20$, 40 and 80. The results for the tests using the ordinary shape-restricted regression estimators are labeled as MREG*

| | Linear regression function | | | | "Ramp" regression function | | |
|---|---|---|---|---|---|---|---|
| $n$ | $F$-test | IQRS B-test | MREG B-test | $n$ | $F$-test | IQRS B-test | MREG B-test |
| 20 | 0.25 | 0.22 | 0.22 | 20 | 0.25 | 0.28 | 0.27 |
| 20 | 0.50 | 0.45 | 0.44 | 20 | 0.50 | 0.63 | 0.61 |
| 20 | 0.75 | 0.69 | 0.68 | 20 | 0.75 | 0.90 | 0.88 |
| 40 | 0.25 | 0.22 | 0.21 | 40 | 0.25 | 0.30 | 0.28 |
| 40 | 0.50 | 0.45 | 0.41 | 40 | 0.50 | 0.62 | 0.59 |
| 40 | 0.75 | 0.70 | 0.66 | 40 | 0.75 | 0.89 | 0.86 |
| 80 | 0.25 | 0.21 | 0.20 | 80 | 0.25 | 0.28 | 0.27 |
| 80 | 0.50 | 0.44 | 0.41 | 80 | 0.50 | 0.60 | 0.56 |
| 80 | 0.75 | 0.69 | 0.65 | 80 | 0.75 | 0.88 | 0.84 |

Table 4

*Power comparisons for the test of linear vs convex regression function. For the spline, the number of interior knots is $l = 2$, 2 and 3 corresponding to $n = 20$, 40 and 80. The results for the tests using the ordinary shape-restricted regression estimators are labeled as CREG*

| | Quadratic regression function | | | | "Ramp" regression function | | |
|---|---|---|---|---|---|---|---|
| $n$ | $F$-test | CQRS B-test | CREG B-test | $n$ | $F$-test | CQRS B-test | CREG B-test |
| 20 | 0.25 | 0.22 | 0.212 | 20 | 0.25 | 0.25 | 0.24 |
| 20 | 0.50 | 0.46 | 0.42 | 20 | 0.50 | 0.54 | 0.52 |
| 20 | 0.75 | 0.70 | 0.66 | 20 | 0.75 | 0.80 | 0.78 |
| 40 | 0.25 | 0.22 | 0.20 | 40 | 0.25 | 0.26 | 0.24 |
| 40 | 0.50 | 0.45 | 0.40 | 40 | 0.50 | 0.53 | 0.49 |
| 40 | 0.75 | 0.70 | 0.64 | 40 | 0.75 | 0.81 | 0.76 |
| 80 | 0.25 | 0.22 | 0.18 | 80 | 0.25 | 0.25 | 0.21 |
| 80 | 0.50 | 0.44 | 0.38 | 80 | 0.50 | 0.54 | 0.49 |
| 80 | 0.75 | 0.69 | 0.62 | 80 | 0.75 | 0.80 | 0.75 |

power than for the ordinary convex regression estimator. For the ramp regression function, the test using the regression splines again has the highest power.

**4. Degrees of freedom and model variance.** Hastie and Tibshirani ([1990](#)) suggest an "effective error degrees of freedom" for a linear smoother: if the estimator is $\hat{\boldsymbol{y}} = \boldsymbol{S}\boldsymbol{y}$, then $df^{err} = n - \mathrm{tr}(2\boldsymbol{S} - \boldsymbol{S}\boldsymbol{S}^T)$. This can be used to estimate the model variance. This definition is consistent with ordinary least-squares regression, where $\boldsymbol{S}$ is a projection matrix and, hence, $2\boldsymbol{S} - \boldsymbol{S}\boldsymbol{S}^T = \boldsymbol{S}$ and $\text{tr}(\boldsymbol{S})$ is the dimension of the linear space defined by the model.

For ordinary shape-restricted regression and for shape-restricted regression splines, the estimator is a mixture of linear estimators, with the mixing distribution corresponding to the probabilities of the data vector falling in the sectors. Let $D$ be $r$ plus the number of edges of the face of the cone on which the projection falls, and let $d$ be the realized value, so that a candidate for error degrees of freedom is $n - d$. Meyer and Woodroofe ([2000](#)) showed that the model variance estimate $SSE/(n - d)$ is too small; in fact,

$$(8) \qquad n - 2E(D) \leq \frac{E(SSE)}{\sigma^2} \leq n - E(D),$$

where $D$ is the random variable indicating the dimension of $J$ plus $r$. They recommend $SSE/(n - 1.5d)$ to estimate the model variance for ordinary monotone regression. Further, for monotone regression, they showed that $E(D) = O_p(n^{1/3})$.

For shape-restricted regression splines, $E(D)$ is limited by the number of knots. If the number of knots grows slowly, as $n^{1/(2p+1)}$ for example, then the model dimension is small compared to the ordinary shape-restricted regression. We may write the SSE as

$$(9) \qquad \|\boldsymbol{y} - \hat{\boldsymbol{\theta}}\|^2 = \left\|\sum_{j \notin J} b_j \boldsymbol{\gamma}^j\right\|^2 + \left\|\sum_{j=1}^{n-m-r} a_j \boldsymbol{w}^j\right\|^2,$$

where the last term may be written as $a_1^2 + \cdots + a_{n-m-r}^2$ if the $\boldsymbol{w}^j$ are chosen to form an orthonormal set. The $a_i$ are mean-zero, independent normal random variables with common variance $\sigma^2$, and because the $\boldsymbol{w}^j$ are orthogonal

TABLE 5

*Comparison of estimators of model standard deviation. The percent bias and standard deviation of four estimators are shown: MQRS = Monotone Quadratic Regression Spline, MQRS (cons) = conservative version, M–W = Meyer–Woodroofe estimator, MLE = Maximum Likelihood Estimator. Results from 10,000 simulated data sets with $f(x) = x^2$; two significant figures presented*

| $n$ | $\sigma$ | MQRS | | MQRS (cons) | | M–W | | MLE | |
|---|---|---|---|---|---|---|---|---|---|
| | | % bias | std dev | % bias | std dev | % bias | std dev | % bias | std dev |
| 20 | 0.1 | −0.66 | 0.0036 | 3.8 | 0.0038 | 26 | 0.0086 | −65 | 0.0019 |
| 40 | 0.1 | −0.43 | 0.0024 | 1.7 | 0.0025 | 8.1 | 0.0035 | −49 | 0.0016 |
| 80 | 0.1 | −0.22 | 0.0016 | 0.55 | 0.0017 | 3.3 | 0.0020 | −35 | 0.0013 |
| 20 | 1.0 | −1.9 | 0.34 | 12 | 0.39 | −5.6 | 0.36 | −28 | 0.27 |
| 40 | 1.0 | −0.77 | 0.24 | 6.8 | 0.25 | −1.5 | 0.25 | −18 | 0.21 |
| 80 | 1.0 | −0.55 | 0.16 | 2.6 | 0.17 | −0.52 | 0.17 | −12 | 0.15 |

to $S$, their distributions are independent of the constraint $\boldsymbol{y} \in \mathcal{C}_J$. Therefore,

$$
(10) \qquad \left\| \sum_{j=1}^{n-m-r} a_j \boldsymbol{w}^j \right\|^2 \Big/ \sigma^2 \sim \chi^2(n-m-r).
$$

We can show that a result similar to (8) holds for the shape-restricted regression splines, so that

$$
m - 2E(D) \leq \frac{\| \sum_{j \notin J} b_j \boldsymbol{\gamma}^j \|^2}{\sigma^2} \leq m - E(D),
$$

where $D$ is the size of $J$. If the shape restrictions are strictly held (i.e., the function is strictly monotone), the constraints are more likely to be unbinding as $n$ gets larger. The value of $D$ for shape-restricted regression splines tends to be closer to $m$, so that $m - 2E(D)$ may be negative. Therefore, we choose as the model variance estimator

$$
(11) \qquad \hat{\sigma}^2 = \frac{SSE}{n-d},
$$

where $d$ is the size of the realized $J$ plus $r$. This estimator tends to have a small negative bias that decreases as $n$ gets larger. An alternative conservative estimator is $\tilde{\sigma}^2 = SSE/(n-m)$; this over-estimates the variance because the denominator is too small.

Simulations show that, for the monotone assumptions, the proposed estimator has smaller bias than the MLE and smaller bias and variance than that using the Meyer–Woodroofe estimator and ordinary monotone regression, for small to moderate sample sizes and several choices of model variance. In Table 5 four estimators of model standard deviation for monotone regression are compared. For each combination of choices of $n$ and $\sigma$, 10,000 datasets were simulated using a linear regression function and i.i.d. normal errors. The MQRS uses increasing quadratic regression spline with variance estimate $SSE/(n-d)$, the MQRS (cons) is $SSE/(n-m)$, and M–W uses standard shape-restricted regression with variance estimate $SSE/(n-1.5d)$. Finally, the MLE uses standard monotone regression and variance estimate $SSE/n$. The percent bias is the difference between the mean of the computed estimator and the true $\sigma$, divided by the true $\sigma$. Simulated estimates of the bias and standard deviation of the estimator of the model variance in convex regression are shown in Table 6, and compared with the MLE.

## APPENDIX

PROOF OF PROPOSITION 4. Given $J \subseteq \{1, \ldots, m\}$, there are four orthogonal linear subspaces of $\mathbb{R}^n$: $V$, $\mathcal{L}(\{\boldsymbol{\delta}^j, j \in J\})$, $\mathcal{L}(\{\boldsymbol{\gamma}^j, j \notin J\})$, and $\mathcal{L}(\{\boldsymbol{w}^j, j = 1, \ldots, n-m-r\})$. Because the dimensions of these subspaces add to $n$, any $\boldsymbol{y} \in \mathbb{R}^n$ can be written as the sum of the projections onto these four subspaces. Let $\hat{\boldsymbol{\theta}} = \boldsymbol{v} + \sum_{j \in J} b_j \boldsymbol{\delta}^j$ be the unique vector in $\mathcal{C}$ that minimizes $\|\boldsymbol{y} - \boldsymbol{\theta}\|^2$ with $b_j > 0$ for $j \in J$. By Proposition 1, we have $\boldsymbol{e} = \boldsymbol{y} - \hat{\boldsymbol{\theta}}$ is in the space spanned by the $\boldsymbol{w}^j$ and the $\boldsymbol{\gamma}^j, j \notin J$, so we write $\boldsymbol{e}$ as a linear combination of these. Now we need only show that the coefficients $b_j$ of the $\boldsymbol{\gamma}^j$ are nonnegative. This is easy by (6) because, for any $j \notin J$, $\langle \boldsymbol{y} - \hat{\boldsymbol{\theta}}, \boldsymbol{\delta}^j \rangle \leq 0$, and

$$\langle \boldsymbol{y} - \hat{\boldsymbol{\theta}}, \boldsymbol{\delta}^j \rangle = \left\langle \sum_{l \notin J} b_l \boldsymbol{\gamma}^l + \sum_{l=1}^{n-m-r} c_l \boldsymbol{w}^l, \boldsymbol{\delta}^l \right\rangle = -b_j. \qquad \square$$

TABLE 6
*Comparison of the estimator of model standard deviation using regression spline for a convex regression function, compared with the maximum likelihood variance estimator. The underlying regression function is $f(x) = x^2$. Results from 10,000 simulated data sets; two significant figures presented*

| | | CQRS | | CQRS (cons) | | MLE | |
|---|---|---|---|---|---|---|---|
| $n$ | $\sigma$ | % bias | std dev | % bias | std dev | % bias | std dev |
| 20 | 0.1 | $-2.8$ | 0.018 | 11. | 0.0039 | $-32$ | 0.0026 |
| 40 | 0.1 | $-1.0$ | 0.012 | 2.9 | 0.0024 | $-18.6$ | 0.0020 |
| 80 | 0.1 | $-0.64$ | 0.0082 | 0.81 | 0.0017 | $-10.9$ | 0.0015 |
| 20 | 1.0 | $-2.4$ | 0.17 | 12.1 | 0.39 | $-20$ | 0.29 |
| 40 | 1.0 | $-0.89$ | 0.12 | 7.1 | 0.25 | $-12$ | 0.21 |
| 80 | 1.0 | 0.58 | 0.081 | 4.1 | 0.17 | $-6.7$ | 0.15 |

## SUPPLEMENTARY MATERIAL

**R code: Supplement 1** (DOI: 10.1214/08-AOAS167SUPPA). Performs a weighted monotone piecewise quadratic spline least-squares regression. inputs: scatterplot points $(x, y)$ where the $x$ are sorted and distinct. weights w must be positive. $\text{var}(y_i) = w_i$  $k$ is the number of interior knots. These will be placed at approximately equal $x$-quantiles output: the values of the fit at the observed $x$ values.

**R code: Supplement 2** (DOI: 10.1214/08-AOAS167SUPPB). Performs a weighted convex piecewise cubic spline least-squares regression. inputs: scatterplot points $(x, y)$ where the $x$ are sorted and distinct. weights $w$ must be positive. $\text{var}(y_i) = w_i$ $k$ is the number of interior knots. These will be placed at approximately equal $x$-quantiles output: the values of the fit at the observed $x$ values.

**R code: Supplement 3** (DOI: 10.1214/08-AOAS167SUPPC). Performs a weighted monotone convex piecewise cubic spline least-squares regression.

inputs: scatterplot points $(x, y)$ where the x are sorted and distinct. weights $w$ must be positive. $\mathrm{var}(y_i) = w_i$ $k$ is the number of interior knots. These will be placed at approximately equal $x$-quantiles output: the values of the fit at the observed $x$ values.

**R code: Supplement 4** (DOI: 10.1214/08-AOAS167SUPPD). Fit two parallel monotone piecewise quadratic curves to a scatterplot. inputs: scatterplot points $(x, d, y)$ where the $x$ are sorted and distinct and $d$ is a vector of ones and zeros. $k$ is the number of interior knots. These will be placed at approximately equal $x$-quantiles outputs: the values of the fit for $d = 0$ at the observed $x$ values and the increase in intercept for $d = 1$.

COLORADO STATE UNIVERSITY
STATISTICS BUILDING
FORT COLLINS, COLORADO 80523-1877
USA
E-MAIL: meyer@stat.colostate.edu